\documentclass[letterpaper, 10 pt, conference]{IEEEtran}  

\IEEEoverridecommandlockouts                              % This command is only needed if 
\usepackage{graphicx}
\usepackage{url}  %IVM added this because of the URLs in the reference list
\usepackage{algorithm}

\usepackage{amsmath}
\usepackage{amsthm}
\usepackage{amssymb}

\usepackage[T1]{fontenc}
\usepackage{gensymb,textcomp}
\usepackage{amsmath}
\usepackage{multirow}
\usepackage{hhline}
\usepackage{booktabs}
\usepackage{siunitx}
\usepackage{array,longtable,multicol,multirow,siunitx,tabularx}
\usepackage{tabularx}
\usepackage[caption=false, font=footnotesize]{subfig}
\usepackage{hyperref}
\usepackage[caption=false,font=footnotesize]{subfig}
\usepackage[backend=biber,style=ieee]{biblatex}
\usepackage{setspace}
\title{\LARGE \bf
BandRouteNet: An Adaptive Band Routing Neural Network\\ for EEG Artifact Removal
}

\author{
        Phat~Lam
\thanks{P. Lam is with Ho Chi Minh City University of Technology, Vietnam.}%
}

\begin{document}

\maketitle
% \thispagestyle{empty}
% \pagestyle{empty}

%%%%%%%%%%%%%%%%%%%%%%%%%%%%%%%%%%%%%%%%%%%%%%%%%%%%%%%%%%%%%%%%%%%%
\begin{abstract}
Electroencephalography (EEG) is highly susceptible to artifact contamination, such as electrooculographic (EOG) and electromyographic (EMG) interference, which severely degrades signal quality and hinders reliable interpretation in applications including neurological diagnosis, brain--computer interfaces (BCIs), etc. Effective EEG denoising remains challenging because different artifact sources exhibit diverse and temporally varying distributions, together with distinct spectral characteristics across frequency bands. To address these issues, we propose \textit{BandRouteNet}, an adaptive frequency-aware neural network for EEG denoising that jointly exploits band-specific processing and full-band contextual modeling. The proposed model performs band-wise denoising to explicitly capture frequency-dependent artifact patterns. Within this framework, we introduce a routing mechanism that adaptively determines where and to what extent denoising should be applied across temporal locations within each frequency band. 
In parallel, a full-band conditioner directly processes the original noisy EEG to extract global temporal context, producing both conditional parameters for modulating the band-wise pathway and a coarse-grained signal-level refinement to supplement the final reconstruction. Extensive experiments on the EEGDenoiseNet benchmark dataset demonstrate that \textit{BandRouteNet} outperforms other methods under EOG, EMG, and mixed-artifact conditions in terms of Relative Root Mean Square Error (RRMSE) and Signal-to-Noise Ratio Improvement (SNR$_{\text{imp}}$) under unified experimental settings, while remaining highly parameter-efficient with only 0.2M trainable parameters. These results highlight its strong potential for high-performance EEG artifact removal in resource-constrained applications.

\end{abstract}

\section{Introduction}

Electroencephalography (EEG) is a non-invasive and widely used modality for monitoring brain activity due to its high temporal resolution, low cost, and portability. It supports a wide range of applications, including brain--computer interfaces (BCIs), neurological monitoring, sleep analysis, and clinical diagnosis~\cite{review_01}. However, EEG recordings are highly vulnerable to physiological artifacts, especially electrooculogram (EOG) and electromyogram (EMG) artifacts, which can distort signal morphology, spectral characteristics, and downstream analysis~\cite{review_02}.

A large body of work has focused on EEG denoising. Many classical methods have been developed, including regression-based filtering\cite{romo2009}, Independent Component Analysis (ICA)~\cite{halder2007, li2021}, Wavelet Transform, and Empirical Mode Decomposition (EMD)~\cite{,kaur2021,hu2022}. Although effective in some scenarios, these approaches often rely on reference channels, multichannel recordings, manual component selection, or sensitive thresholding procedures. More recently, deep learning methods have shown strong potential by learning direct mappings from noisy EEG to clean EEG. Zhang et al.~\cite{eegdenoisenet} established  end-to-end baselines using fully connected networks (FCNNs), convolutional neural networks (CNNs), and recurrent neural networks (RNNs). Subsequent studies explored more advanced architectures, including the transformer-based EEGDNet~\cite{EEGDnet}, the embedding-separation framework DeepSeparator~\cite{deepseparator}, the U-Net-based intepretable archiecture LRR-Unet~\cite{lrr_unet}, GAN-based approaches for artifact suppression~\cite{gan_1,gan_2}, etc. Although these methods have shown strong denoising performances, several limitations has remained, suggesting rooms for prospective improvements.

First, different artifact sources exhibit markedly different spectral characteristics (e.g. EOG are typically concentrated in low-frequency bands , whereas EMG are broaderband and often more prominent in higher frequencies)~\cite{eog_ref,emg_ref}. However, frequency-aware artifact modeling is still insufficiently explored~\cite{eeg_denoising_survey}. 
Second, many existing denoising models apply relatively uniform processing across the signal~\cite{eeg_denoising_survey}. despite the fact that EEG artifacts are often non-stationary and temporally localized. Such uniform treatment may over-smooth clean EEG regions while failing to adequately suppress severely corrupted segments. This motivates an adaptive denoising method that jointly captures frequency-specific artifact patterns and time-varying contamination, enabling selective artifact suppression while preserving underlying neural activity.

To address these issues, we propose \textit{BandRouteNet}, a frequency-aware and temporally adaptive EEG denoising network. The proposed model consists of two cooperative components: a \textit{Band-specific Denoiser}, which operates on decomposed EEG frequency bands to suppress artifacts in a band-aware manner, and a \textit{Full-band Conditioner}, which processes the original fullband signal to provide global temporal guidance for band-wise denoising. In addition, an \textit{artifact routing mechanism} is introduced to adaptively modulate denoising strength over time and frequency bands, allowing the model to focus more strongly on contaminated band--time regions while preserving cleaner neural activity.
\begin{figure*}[t]
	\centering
	
	\subfloat[Overall framework of \textit{BandRouteNet}.\label{fig:overall-architecture}]{
		\includegraphics[width=0.55\linewidth, ]{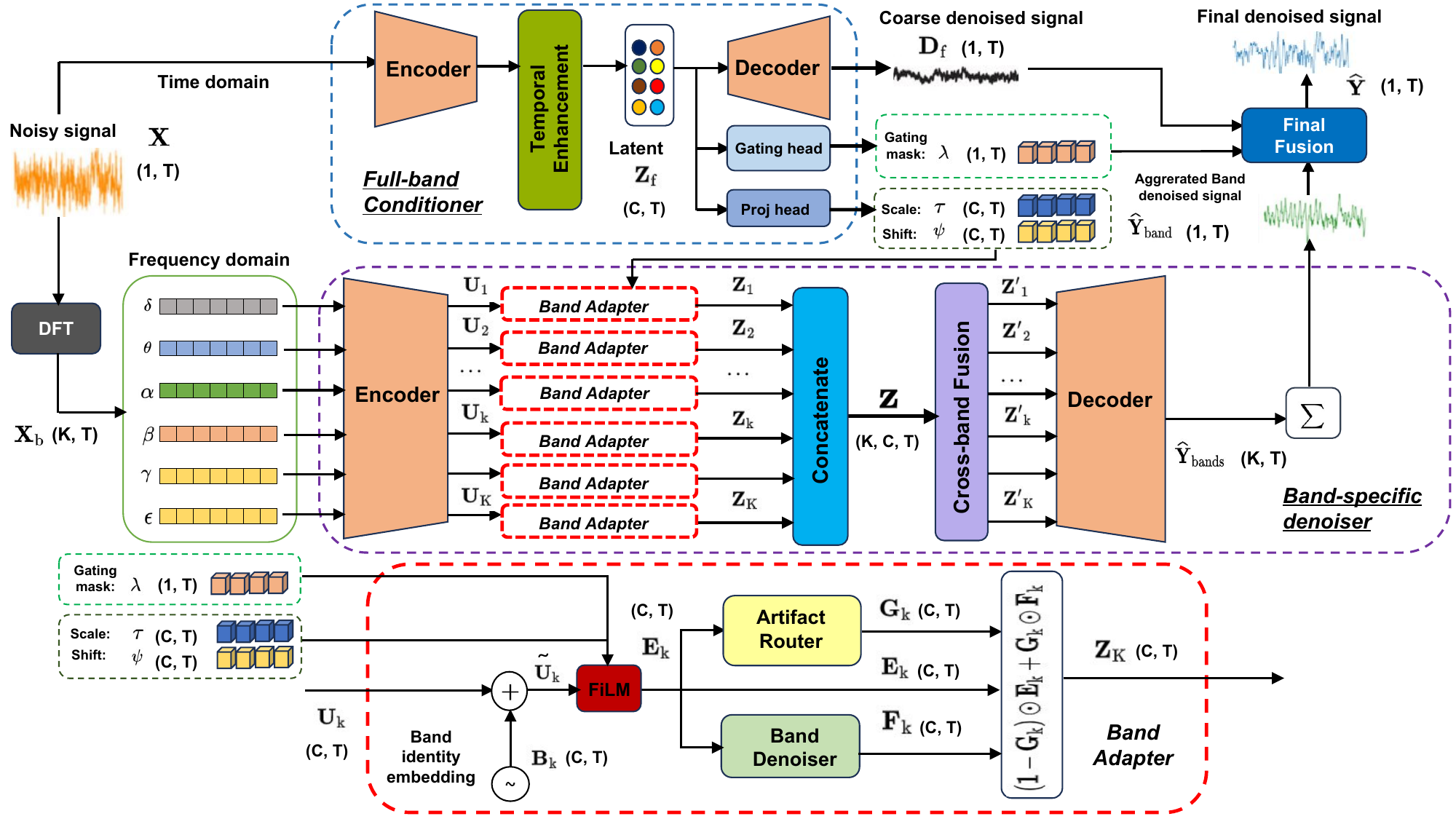}
	}
	\hfill
	\subfloat[Detailed structures of the main component blocks.\label{fig:overall-v2}]{
		\includegraphics[width=0.42\linewidth, height = 0.32\linewidth]{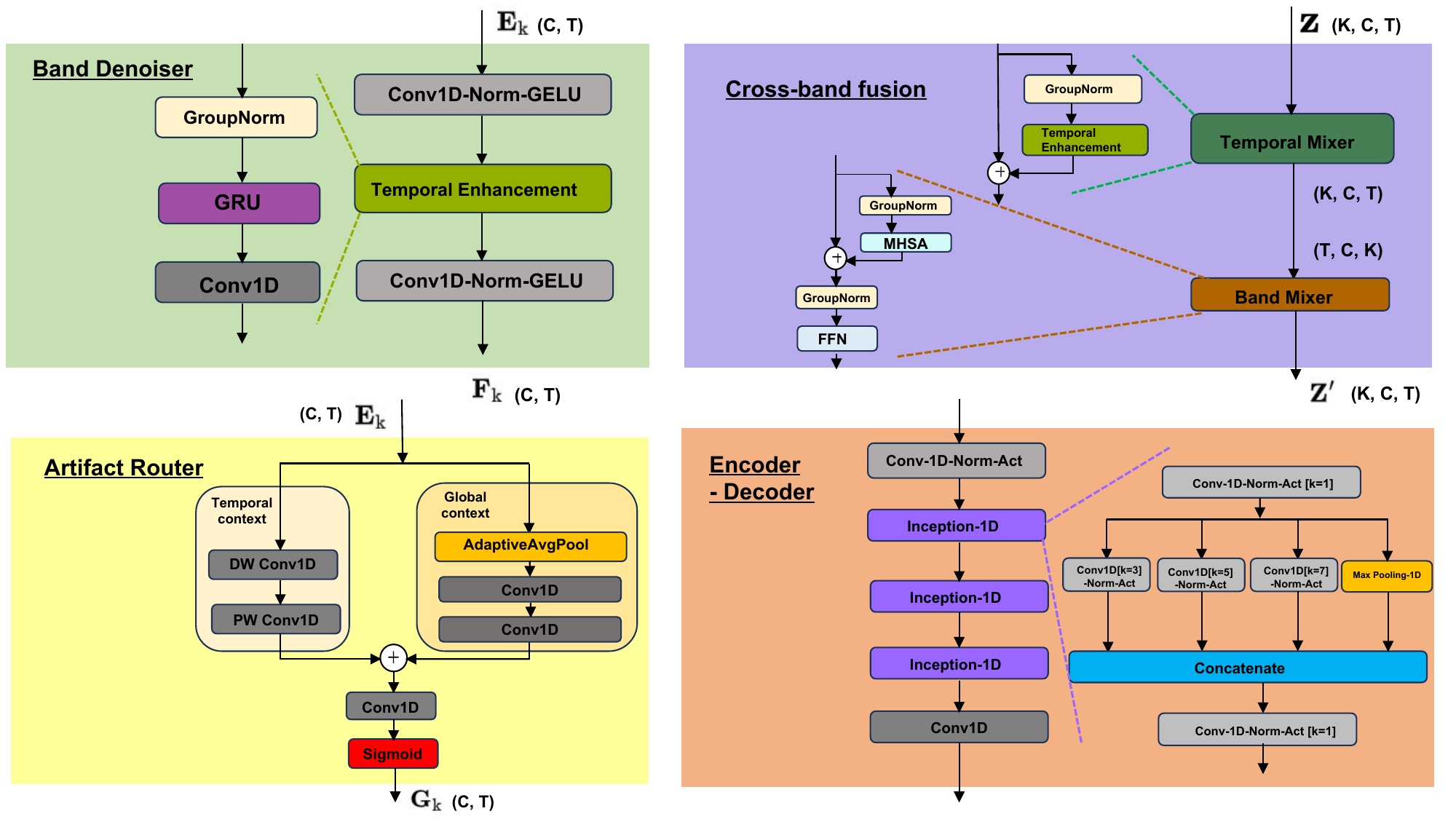}
	}
	
	\caption{Overview of the proposed \textit{BandRouteNet} architecture for EEG artifact removal}
	\label{fig:overall}
	\vspace{-0.55cm}
\end{figure*}
The main contributions of this work are summarized as follows:
\begin{itemize}
	\item We propose \textit{BandRouteNet}, a frequency-aware EEG denoising architecture that combines band-specific processing, supplemented by full-band temporal conditioning.
	\item We introduce an adaptive artifact routing mechanism that learns time-varying denoising emphasis across latent band features, improving selective artifact suppression for non-stationary EEG contamination.
	\item Through extensive experiments on the EEGDenoiseNet benchmark, we show that the proposed method achieves superior denoising performance over other methods across EOG, EMG, and mixed-noise conditions under unified evaluation settings, while remaining highly parameter-efficient.
\end{itemize}
\section{Method}

\subsection{Overall Architecture}

The proposed \textit{BandRouteNet} is a dual-path EEG denoising framework that combines frequency-aware band-wise processing with full-band temporal conditioning. Given a noisy EEG segment $\mathbf{X}\in\mathbb{R}^{1\times T}$, the model estimates the clean signal $\hat{\mathbf{Y}}\in\mathbb{R}^{1\times T}$ through two cooperative branches: a \textit{Band-specific Denoiser} and a \textit{Full-band Conditioner}, as illustrated in Fig.~\ref{fig:overall-architecture}.

The Band-specific Denoiser decomposes the noisy signal into multiple frequency bands and performs adaptive denoising within each band. This allows the model to handle frequency-dependent artifact characteristics. In parallel, the Full-band Conditioner operates on the original full-band signal to provide a coarse signal-level refinement and generate global contextual modulation parameters for guiding the band-wise pathway. The final denoised output is obtained by adaptively fusing the reconstructed band-wise signal with the full-band refinement.

\subsection{Full-band Conditioner}
\label{full-band-conditioner}

The Full-band Conditioner models the original noisy signal to provide global temporal guidance for the band-wise branch. The input signal is first encoded into a latent representation:
\begin{equation}
	\mathbf{H}_f = \mathcal{E}_f(\mathbf{X}), \qquad \mathbf{H}_f\in\mathbb{R}^{C\times T},
\end{equation}
where $\mathcal{E}_f(\cdot)$ denotes the full-band encoder. The encoded feature is then enhanced by a temporal modeling block:
\begin{equation}
	\mathbf{Z}_f = \mathcal{T}_f(\mathbf{H}_f),
\end{equation}
where $\mathcal{T}_f(\cdot)$ is implemented using Group Normalization, GRU-based temporal modeling, and 1D convolution.

From $\mathbf{Z}_f$, the conditioner produces three outputs. First, a decoder generates a coarse full-band denoised signal:
\begin{equation}
	\mathbf{D}_f = \mathcal{D}_f(\mathbf{Z}_f), \qquad \mathbf{D}_f\in\mathbb{R}^{1\times T}.
\end{equation}
Second, a gating head predicts a temporal fusion weight leveraged for the final reconstruction:
\begin{equation}
	\boldsymbol{\lambda} = \sigma(\mathcal{H}_{\lambda}(\mathbf{Z}_f)), \qquad
	\boldsymbol{\lambda}\in\mathbb{R}^{1\times T},
\end{equation}
where $\sigma(\cdot)$ denotes the sigmoid activation. Third, a 1D convolution-based projection head generates feature modulation parameters:
\begin{equation}
	[\boldsymbol{\tau}, \boldsymbol{\psi}] = \mathcal{H}_{\mathrm{proj}}(\mathbf{Z}_f),
	\qquad
	\boldsymbol{\tau},\boldsymbol{\psi}\in\mathbb{R}^{C\times T}.
\end{equation}
These parameters are used to condition the band-specific latent features through feature-wise modulation.

\subsection{Band-specific Denoiser}
\label{band-specific-denoiser}

The Band-specific Denoiser exploits the spectral heterogeneity of EEG artifacts by processing decomposed frequency bands separately. The noisy signal is first transformed into the frequency domain using the Discrete Fourier Transform (DFT):
\begin{equation}
	\hat{x}[f] = \sum_{t=0}^{T-1} x[t] e^{-j2\pi ft/T}.
\end{equation}
The spectrum is divided into $K$ frequency bands using band masks $\{\mathbf{M}_k\}_{k=1}^{K}$, and each band signal is reconstructed by inverse DFT:
\begin{equation}
	\mathbf{x}_k = \mathrm{IDFT}(\mathbf{M}_k \odot \hat{\mathbf{x}}),
	\qquad k=1,\ldots,K.
\end{equation}
The resulting band representation is denoted as
\begin{equation}
	\mathbf{X}_b = \{\mathbf{x}_1,\mathbf{x}_2,\ldots,\mathbf{x}_K\}\in\mathbb{R}^{K\times T}.
\end{equation}
Next, each band $\mathbf{x}_k$ is encoded by a shared band encoder:
\begin{equation}
	\mathbf{U}_k = \mathcal{E}_b(\mathbf{x}_k), \qquad
	\mathbf{U}_k\in\mathbb{R}^{C\times T}.
\end{equation}
To distinguish different bands in the shared latent space, a learnable band identity embedding $\mathbf{B}_k$ is added:
\begin{equation}
	\tilde{\mathbf{U}}_k = \mathbf{U}_k + \mathbf{B}_k.
\end{equation}

\begin{table*}[t]
	\centering
	\caption{Performance Comparison of Average Performances across All SNR Levels. The smaller $\mathrm{RRMSE}_\text{t}$, $\mathrm{RRMSE}_\text{s}$, and
		the Larger $\mathrm{CC}$, $\mathrm{SNR}_{\text{imp}}$ the Better Denoising Effect. Best Results are in Bold; Second-best Ones are Underlined}
	\label{tab:benchmark_results}
	\setlength{\tabcolsep}{4pt}
	\renewcommand{\arraystretch}{1.05}
	\resizebox{\textwidth}{!}{%
		\begin{tabular}{l c cccc cccc cccc}
			\hline
			\multirow{2}{*}{\textbf{Method}} & \multirow{2}{*}{\textbf{Parameter (M)}} 
			& \multicolumn{4}{c}{\textbf{EOG dataset}} 
			& \multicolumn{4}{c}{\textbf{EMG dataset}} 
			& \multicolumn{4}{c}{\textbf{Mixed EOG/EMG dataset}} \\
			% \cline{3-6} \cline{7-10} \cline{11-14}
			\cmidrule(lr){3-6} \cmidrule(lr){7-10} \cmidrule(lr){11-14} 
			
			& 
			& \textbf{RRMSE$_{\text{t}}$} & \textbf{RRMSE$_{\text{s}}$} & \textbf{CC} & \textbf{SNR{$_\text{imp}$}}
			& \textbf{RRMSE$_{\text{t}}$} & \textbf{RRMSE$_{\text{s}}$} & \textbf{CC} & \textbf{SNR{$_\text{imp}$}}
			& \textbf{RRMSE$_{\text{t}}$} & \textbf{RRMSE$_{\text{s}}$} & \textbf{CC} & \textbf{SNR{$_\text{imp}$}} \\
			\hline
			FCNN~\cite{eegdenoisenet}            & 1.051 & 0.5570 & 0.5865 & 0.8111 & 10.8011 & \underline{0.6176} & 0.6687 & \textbf{0.7849} & \underline{9.6891} & \underline{0.6887} & 0.7362 & \underline{0.7123} & \underline{11.7388} \\
			Simple-CNN~\cite{eegdenoisenet}     & 16.82 & 0.4437 & 0.4432 & 0.8847 & 12.6351 & 0.7215 & 0.7263 & 0.7255 & 8.4879 & 0.7787 & 0.7258 & 0.6763 & 10.7515 \\
			1D-ResCNN~\cite{1D-ResCNN}     & 8.46  & \underline{0.4184} & \underline{0.4046} & 0.8966 & \underline{13.1761} & 0.7510 & 0.7093 & 0.7076 & 8.2550 & 0.8674 & 0.9065 & 0.6625 & 10.0200 \\
			RNN-LSTM~\cite{eegdenoisenet}        & 0.788 & 0.6903 & 0.7231 & 0.7176 & 8.6099 & 0.6758 & 0.7215 & 0.7384 & 8.8269 & 0.7522 & 0.8123 & 0.6644 & 10.8124 \\
			EEGDnet~\cite{EEGDnet}  & 0.895 & 0.4406 & 0.4125 & 0.8907 & 12.4712 & 0.6554 & 0.6196 & 0.7456 & 9.0505 & 0.6897 & \underline{0.6210} & 0.7125 & 11.5704 \\
			Deep Separator~\cite{deepseparator}  & \textbf{0.032} & 0.4944 & 0.5484 & 0.8658 & 11.3920 & 0.6982 & 0.6899 & 0.7396 & 8.6998 & 0.7593 & 0.7174 & 0.6814 & 10.7386 \\
			LRR-UNet~\cite{lrr_unet}        & 3.19  & 0.4322 & 0.4311 & \underline{0.8967} & 12.7577 & 0.6415 & \underline{0.5827} & 0.7564 & 9.3746 & 0.7144 & 0.6328 & 0.7068 & 11.3037 \\
			\hline
			\textbf{BandRouteNet (Ours)} & \underline{0.20} 
			& \textbf{0.3831} & \textbf{0.3797} & \textbf{0.9156} & \textbf{13.9819}
			& \textbf{0.5962} & \textbf{0.5276} & \underline{0.7802} & \textbf{10.0041}
			& \textbf{0.6605} & \textbf{0.5998} & \textbf{0.7359} & \textbf{11.9310} \\
			\hline
		\end{tabular}%
	}
	\vspace{-0.55cm}
\end{table*}
\begin{table}[t]
	\centering
	\caption{Ablation Study Results on EOG Dataset}
	\label{tab:ablation}
	\renewcommand{\arraystretch}{1.0}
	\scalebox{0.75}{%
		\begin{tabular}{lcccc}
			\hline
			\textbf{Experiments} & \textbf{RRMSE$_{\text{t}}$} & \textbf{RRMSE$_{\text{s}}$} & \textbf{CC} & \textbf{SNR{$_\text{imp}$}} \\
			\hline
			Full model                        & \textbf{0.3831} & \textbf{0.3797} & \textbf{0.9156} & \textbf{13.9819} \\
			W/o Full-band Conditioner          & 0.3907 & 0.3979 & 0.9106 & 13.2324 \\
			W/o Artifact Router   & 0.4091 & 0.4013 & 0.9112 & 13.7610 \\
			W/o Cross-band Fusion             & 0.3956 & 0.4009 & 0.9139 & 13.8310 \\
			W/o Band identity embedding       & 0.3867 & 0.3998 & 0.9130 & 13.8392 \\
			\hline
		\end{tabular}%
	}
	\vspace{-0.6cm}
\end{table}
%\begin{figure}[h]
%	\centering
%	
%	\scalebox{0.85}{%
%		\begin{minipage}{\linewidth}
%			\centering
%			
%			% Row 1
%			\subfloat[RRMSE temporal \label{fig:sub1}]{
%				\includegraphics[width=0.47\linewidth]{figures/eog_rmsse_temporal.pdf}
%			}
%			\hfill
%			\subfloat[RRMSE spectral\label{fig:sub2}]{
%				\includegraphics[width=0.47\linewidth]{figures/eog_rmsse_spectral.pdf}
%			}
%			
%			\vspace{0.05cm}
%			
%			\subfloat[Correlation coefficient\label{fig:sub3}]{
%				\includegraphics[width=0.47\linewidth]{figures/eog_cc.pdf}
%			}
%			\hfill
%			\subfloat[SNR improvement\label{fig:sub4}]{
%				\includegraphics[width=0.47\linewidth]{figures/eog_snr_improvement.pdf}
%			}
%		\end{minipage}%
%	}
%	
%	\caption{Performance Metrics on different SNR Levels (EOG Dataset)}
%	\label{fig:eog-multi-snr}
%\end{figure}
The full-band modulation parameters $(\boldsymbol{\tau},\boldsymbol{\psi})$ are then applied to the band feature using Feature-wise Linear Modulation (FiLM)~\cite{film}:
\begin{equation}
	\mathbf{E}_k =
	\tilde{\mathbf{U}}_k \odot \left(1+\tanh(\boldsymbol{\tau})\right)
	+ \boldsymbol{\psi},
\end{equation}
where $\mathbf{E}_k$ denotes the globally conditioned band feature.

\subsection{Artifact Routing and Band Adaptation}
\label{routing-mechanism}

For each conditioned band feature $\mathbf{E}_k$, the Band Adapter contains two parallel modules: an Artifact Router and a Band Denoiser. The Artifact Router estimates a soft routing mask:
\begin{equation}
	\mathbf{G}_k = \mathcal{R}(\mathbf{E}_k), \qquad
	\mathbf{G}_k\in[0,1]^{C\times T},
\end{equation}
where $\mathbf{G}_k$ controls the denoising strength at each temporal location and channel. The Band Denoiser generates a candidate refined feature:
\begin{equation}
	\mathbf{F}_k = \mathcal{B}(\mathbf{E}_k), \qquad
	\mathbf{F}_k\in\mathbb{R}^{C\times T}.
\end{equation}

The final routed feature is obtained by adaptive interpolation between the original conditioned feature and the denoised proposal:
\begin{equation}
	\mathbf{Z}_k =
	(1-\mathbf{G}_k)\odot \mathbf{E}_k
	+
	\mathbf{G}_k\odot \mathbf{F}_k.
\end{equation}
Thus, small values of $\mathbf{G}_k$ preserve the original representation, while larger values apply stronger denoising. This routing mechanism enables adaptive artifact suppression across both time and frequency.

\subsection{Cross-band Fusion and Final Reconstruction}

After routing, the band features are stacked as $\mathbf{Z}\in\mathbb{R}^{K\times C\times T}.$ A cross-band fusion module is then applied to model dependencies among frequency bands. It first performs temporal mixing within each band and then applies band mixer with Multi-head Self-attention (MHSA)~\cite{attention} to exchange information across bands:
\begin{equation}
	\mathbf{Z}^{(t)} = \mathcal{M}_{\mathrm{temp}}(\mathbf{Z}), 
	\qquad
	\mathbf{Z}' = \mathcal{M}_{\mathrm{band}}(\mathbf{Z}^{(t)}).
\end{equation}

Each refined band feature is decoded into a denoised band signal:
\begin{equation}
	\hat{\mathbf{Y}}_k = \mathcal{D}_b(\mathbf{Z}'_k),
	\qquad
	\hat{\mathbf{Y}}_k\in\mathbb{R}^{1\times T}.
\end{equation}
The band-wise reconstruction $\hat{\mathbf{Y}}_{\mathrm{band}}$ is obtained by summing all decoded bands.
Finally, the band-wise reconstruction is fused with the coarse full-band denoised signal:
\begin{equation}
	\hat{\mathbf{Y}} =
	\hat{\mathbf{Y}}_{\mathrm{band}}
	+
	\boldsymbol{\lambda}\odot \mathbf{D}_f,
\end{equation}
where $\boldsymbol{\lambda}$ adaptively controls the contribution of the full-band refinement. This final fusion combines fine-grained frequency-specific denoising with global signal-level correction.

\begin{figure*}[t]
	\centering
	
	\subfloat[EOG-contaminated segment\label{fig:denoise_EOG}]{
		\includegraphics[width=0.315\linewidth, height=0.22\linewidth]{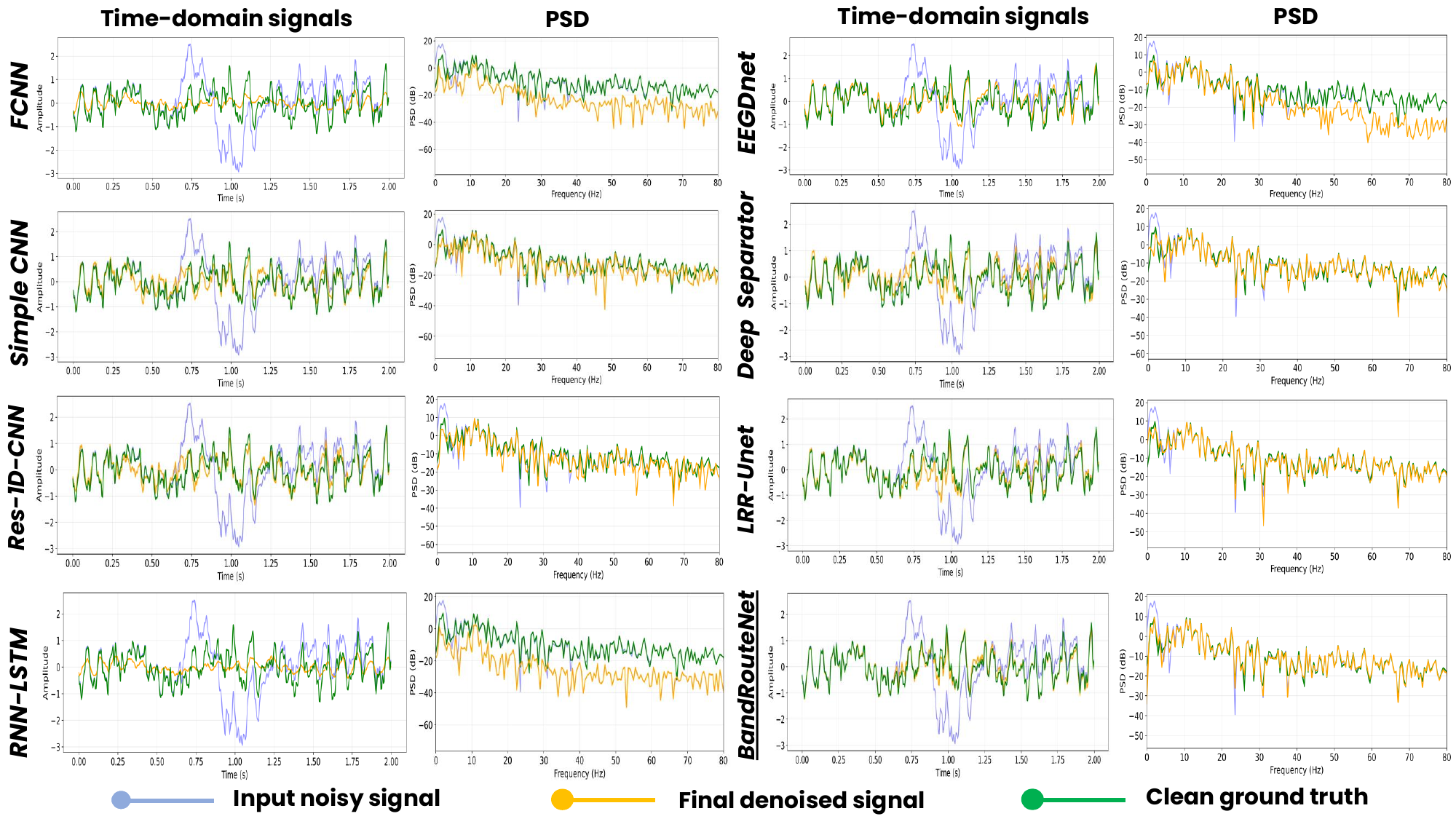}
	}
	\hfill
	\subfloat[EMG-contaminated segment\label{fig:denoise_EMG}]{
		\includegraphics[width=0.315\linewidth, height=0.22\linewidth]{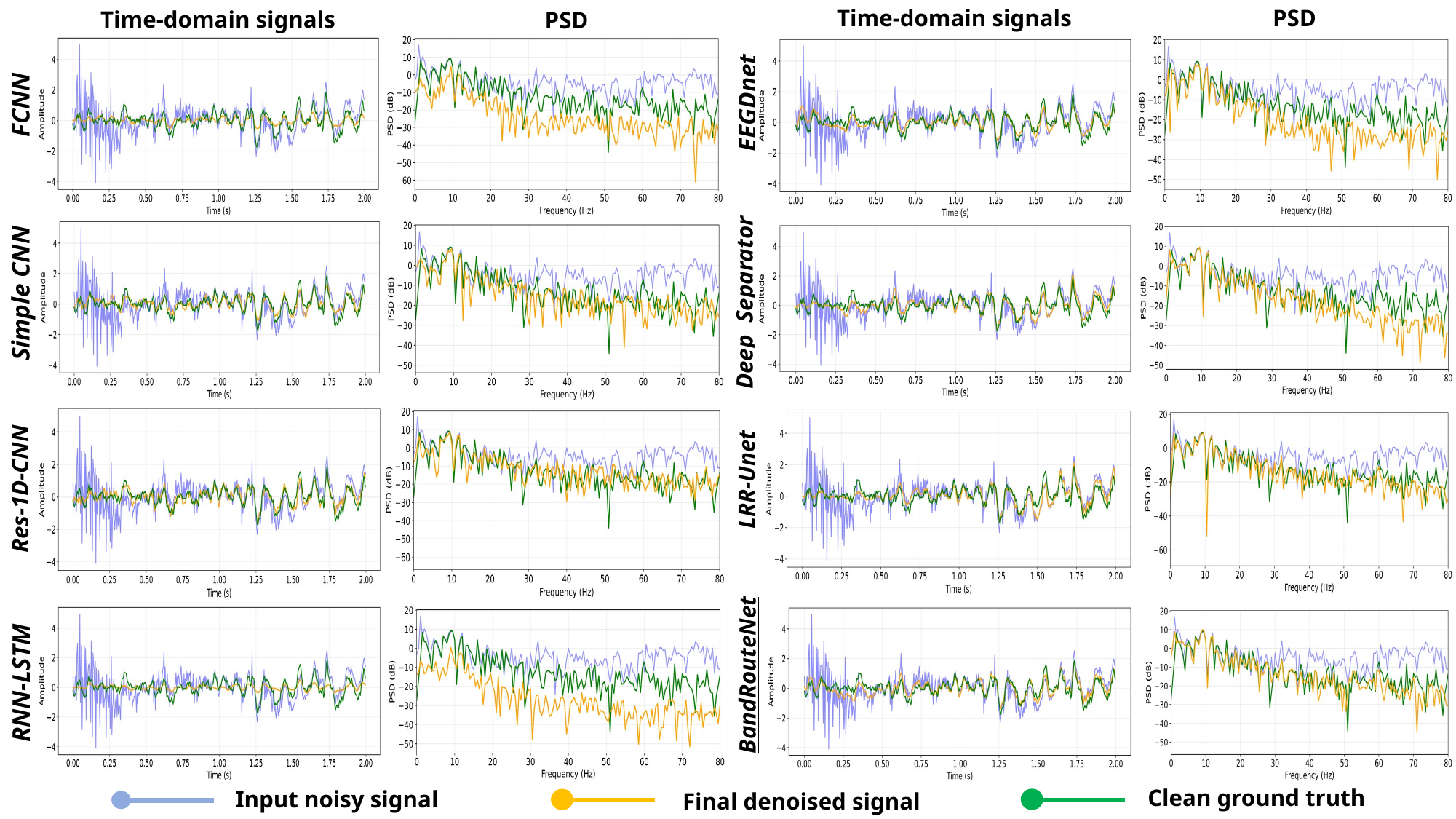}
	}
	\hfill
	\subfloat[Mixed-noise segment\label{fig:denoise_mix}]{
		\includegraphics[width=0.315\linewidth, height=0.22\linewidth]{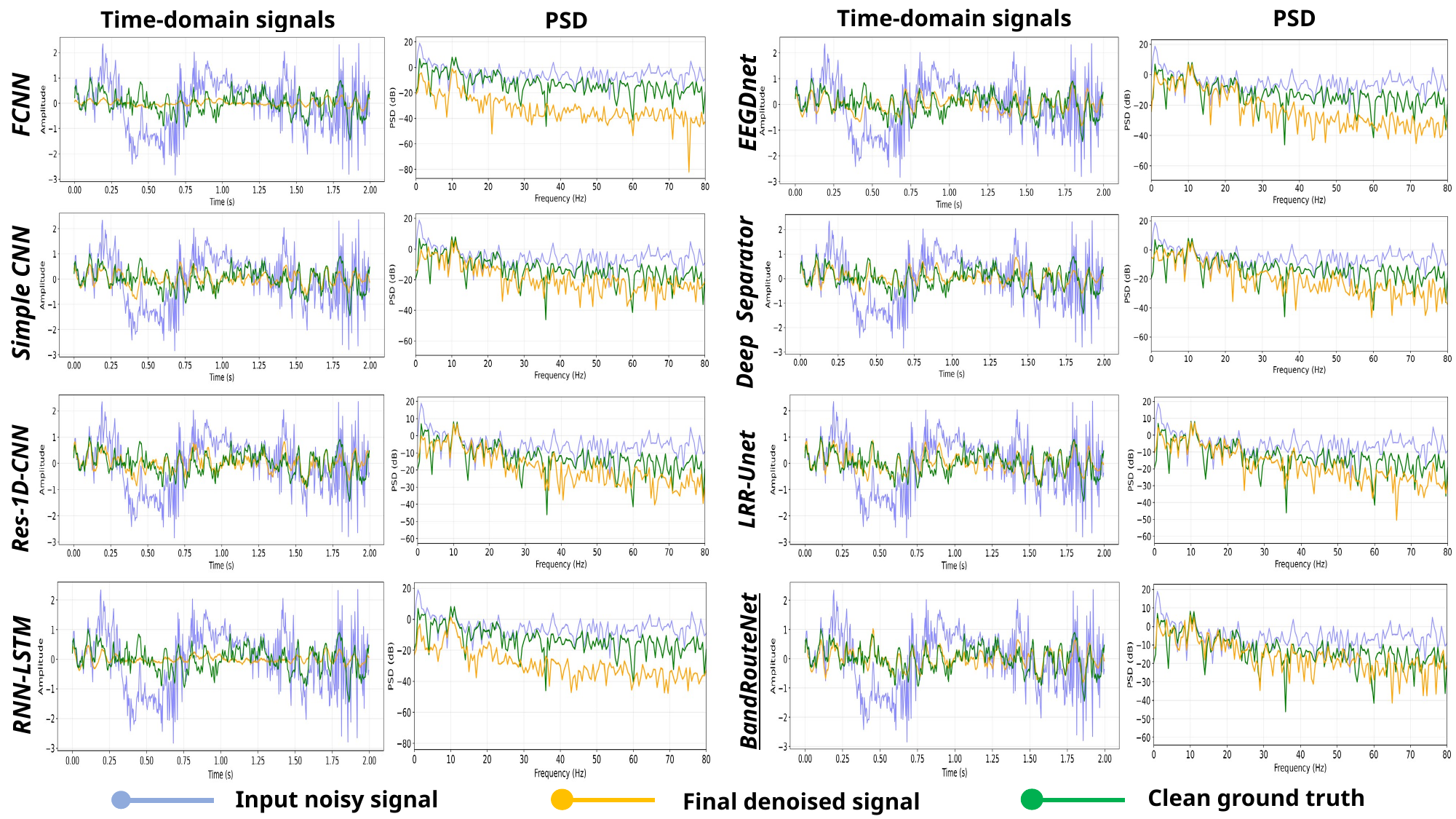}
	}
	
	\caption{Denoising performance comparison on different contaminated EEG segments}
	\label{fig:denoise_comparison}
	\vspace{-0.55cm}
\end{figure*}
\begin{figure}[h]
	\centering
	
	\scalebox{0.9}{%
		\begin{minipage}{\linewidth}
			\centering
			
			% Row 1
			\subfloat[RRMSE temporal \label{fig:sub1}]{
				\includegraphics[width=0.47\linewidth]{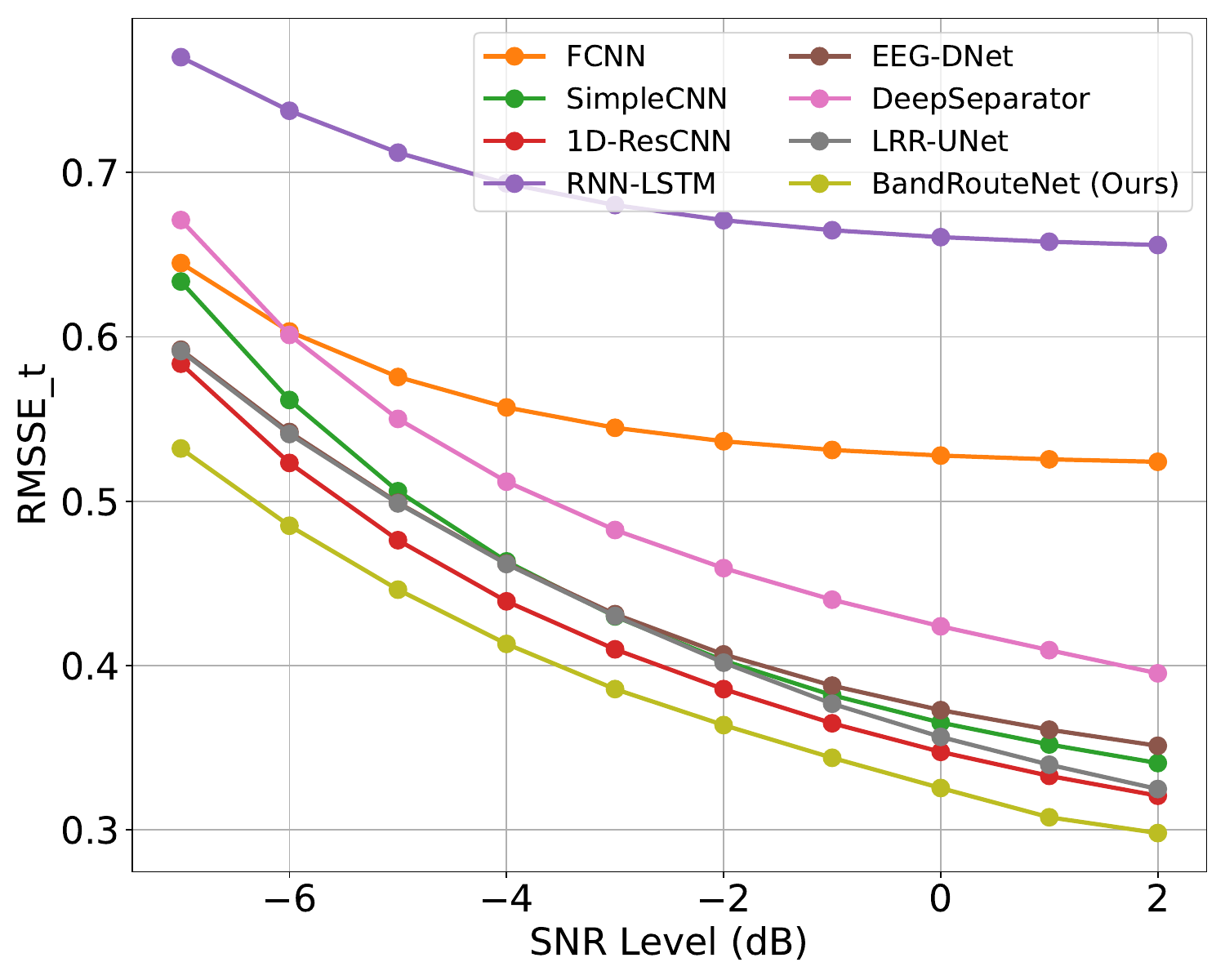}
			}
			\hfill
			\subfloat[RRMSE spectral\label{fig:sub2}]{
				\includegraphics[width=0.47\linewidth]{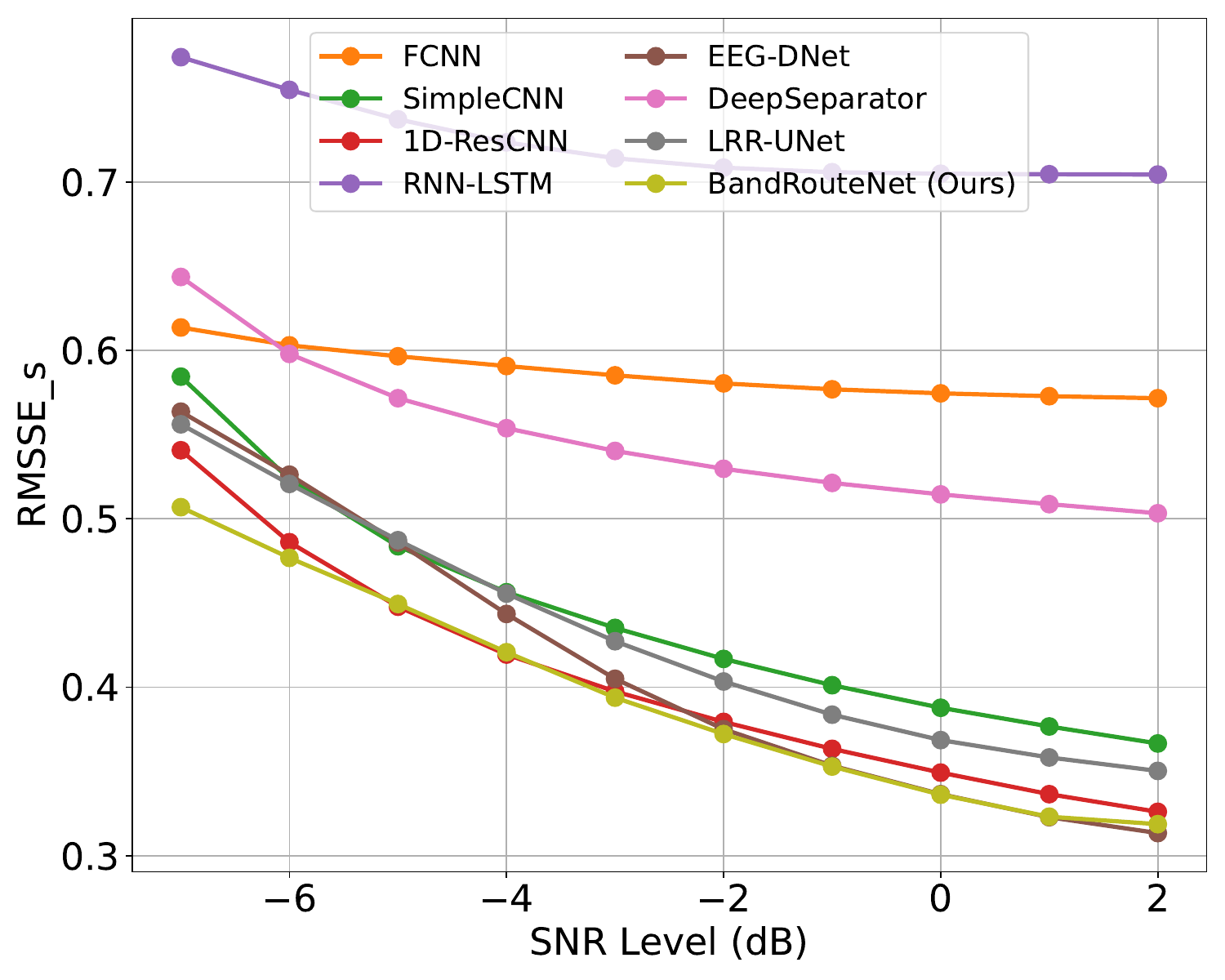}
			}
			
			\vspace{0.05cm}
			
			\subfloat[Correlation coefficient\label{fig:sub3}]{
				\includegraphics[width=0.47\linewidth]{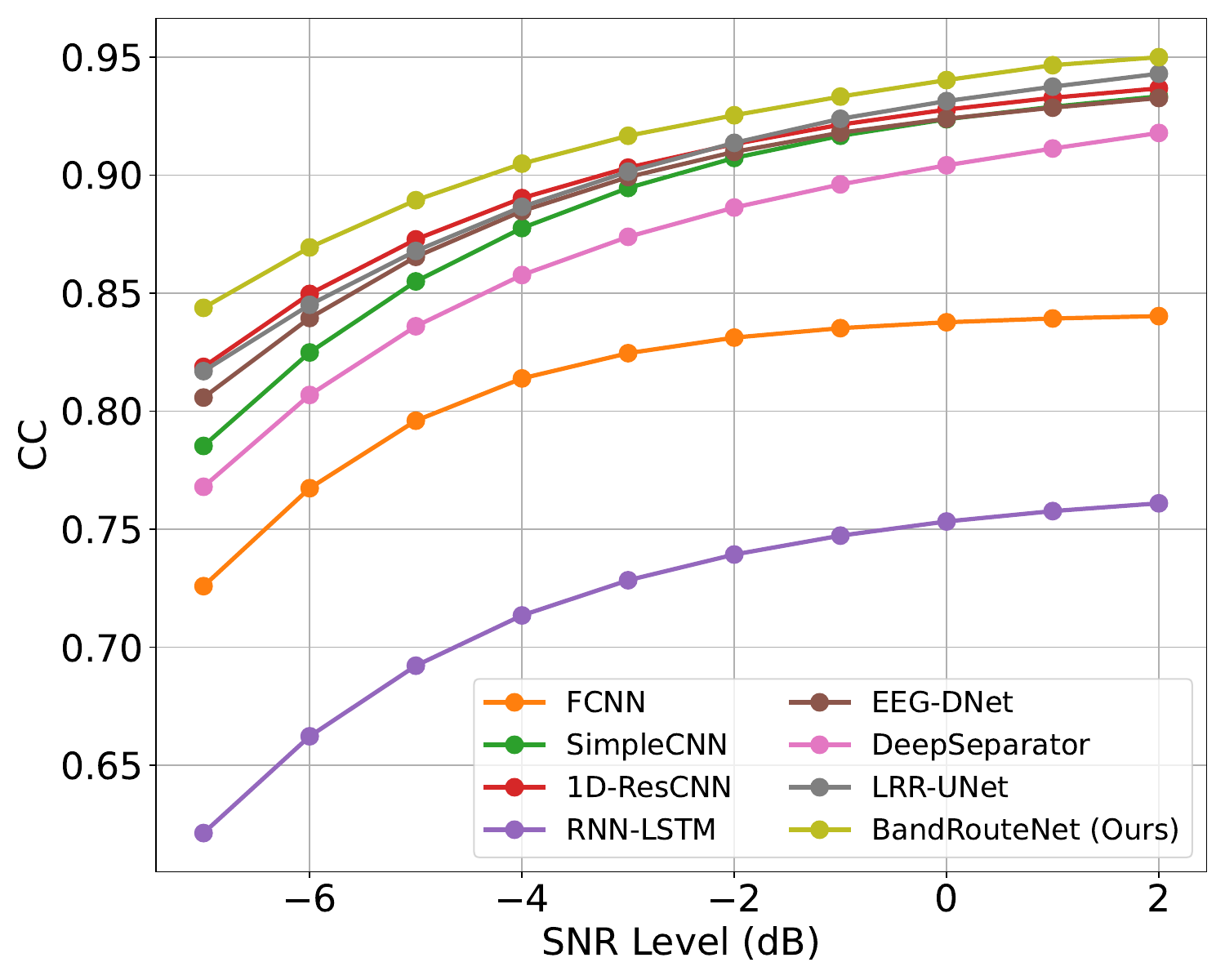}
			}
			\hfill
			\subfloat[SNR improvement\label{fig:sub4}]{
				\includegraphics[width=0.47\linewidth]{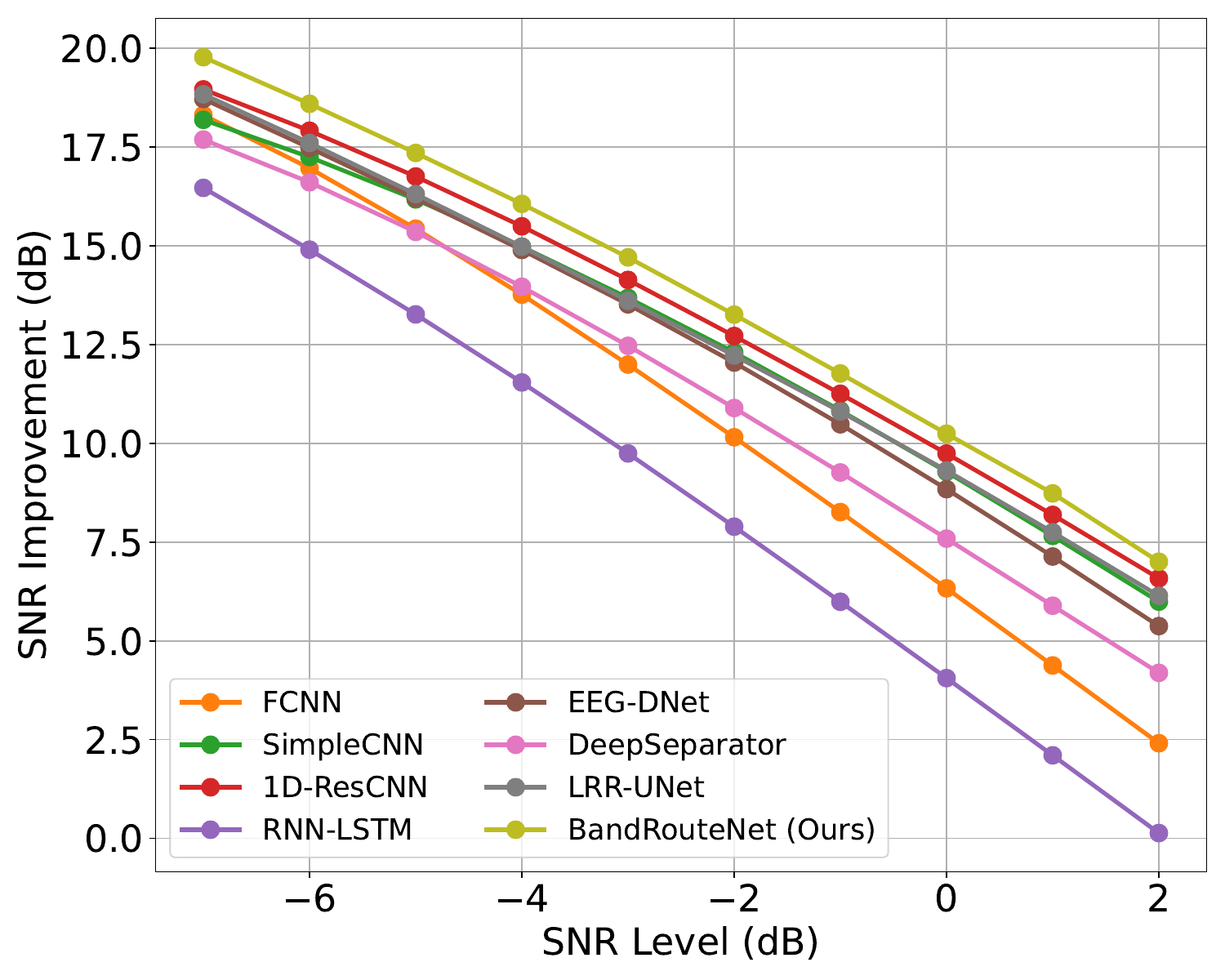}
			}
		\end{minipage}%
	}
	
	\caption{Performance Metrics on different SNR Levels (EOG Dataset)}
	\label{fig:eog-multi-snr}
	\vspace{-0.6cm}
\end{figure}
\section{Experimental Settings}
\subsection{Datasets}

To evaluate the denoising performance of the proposed
model, we adopt the EEGDenoiseNet benchmark ~\cite{eegdenoisenet}. The dataset contains 4515 clean EEG segments, 3400 electrooculogram (EOG) artifact segments, and 5598 electromyogram (EMG) artifact segments. Each segment is 2 s long and sampled at 256 Hz.

Given a clean EEG signal $x$ and an artifact signal $N$, the contaminated EEG signal $y$ is generated by linear mixing:
\begin{equation}
	y = x + \lambda N,
\end{equation}
where $\lambda$ controls the artifact intensity according to the target signal-to-noise ratio (SNR):
\begin{equation}
	\mathrm{SNR} = 10 \log \frac{\mathrm{RMS}(x)}{\mathrm{RMS}(\lambda N)},
\end{equation}
with $\mathrm{RMS}$ denotes the Root Mean Square~\cite{eegdenoisenet}
%with
%\begin{equation}
%	\mathrm{RMS}(g) = \sqrt{\frac{1}{N}\sum_{i=1}^{N} g_i^2}.
%\end{equation}

In this study, three contaminated datasets are constructed: EOG-contaminated EEG, EMG-contaminated EEG, and mixed EOG/EMG-contaminated EEG. For the EOG dataset, 3400 EEG segments are paired with 3400 EOG artifact segments. For the EMG dataset, EEG segments are randomly reused to match the 5598 EMG artifact segments. For the mixed-artifact dataset, EEG, EOG, and EMG segments are aligned to the same sample size, and the noisy signal is generated as:
\begin{equation}
	y = x + \lambda (N_{\mathrm{EOG}} + N_{\mathrm{EMG}}).
\end{equation}

We follow the official EEGDenoiseNet protocol~\cite{eegdenoisenet} for data generation and preprocessing. Specifically, each EEG--artifact pair is augmented across 10 SNR levels, ranging from $-7$ to $2$ dB. The resulting dataset is split into training, validation, and test sets using an 8:1:1 ratio.

\subsection{Evaluation Metrics}
\label{evaluation_metrics}
 We adopt the standard evaluation metrics from benchmarks, including the Temporal Root Mean Squared Error (RRMSE$_{\mathrm{t}}$), Spectral Root Mean Squared Error (RRMSE$_{\mathrm{s}}$), and Correlation Coefficient (CC)~\cite{eegdenoisenet}. In addition, we also report the Signal-to-noise Ratio Improvement (SNR$_{\mathrm{imp}}$) to measure the gain in signal quality before and after denoising.

\subsection{Implementation Details}

The proposed model was implemented in PyTorch and trained on an NVIDIA Tesla T4 GPU. We used the AdamW optimizer~\cite{adamw} with an initial learning rate of $1\times10^{-3}$ and a weight decay of $1\times10^{-4}$. The model was trained for 15 epochs using Mean Squared Error (MSE) as the loss function.

For model configuration, the input noisy EEG signal was decomposed into $K=6$ frequency bands: Delta ($\delta$, 0–-4 Hz), Theta ($\theta$, 4–-8 Hz), Alpha ($\alpha$, 8–-12 Hz), Beta ($\beta$, 13–-30 Hz), Gamma ($\gamma$, 30–-80 Hz), and a higher band $\epsilon$ (80–-128 Hz). The batch size, the hidden feature dimension $C$, and the signal length $T$ are set to 32, 64, 512, respectively.

Both the \textit{Band-specific Denoiser} and the \textit{Full-band Conditioner} adopt an Inception-based encoder–-decoder architecture to capture multi-scale features, as shown at the Figure~\ref{fig:overall-v2}. Each module consists of two encoder stages and two decoder stages, with each stage incorporating two Inception1D blocks. 
% The detailed implementation can be found at: \hyperlink{https://github.com/LTPhat/BandRouteNet}{https://github.com/LTPhat/BandRouteNet}

%\begin{figure}
%    \centering
%    \includegraphics[width=0.97\linewidth, height = 0.4\linewidth]{figures/BRN_EOG.pdf}
%    \caption{Denoising performance comparison on the EOG-contaminated segment}
%    \label{fig:denoise_EOG}
%\end{figure}
%
%
%\begin{figure}
%    \centering
%    \includegraphics[width=0.97\linewidth, height = 0.4\linewidth]{figures/BRN_EMG}
%    \caption{Denoising performance comparison on the EMG-contaminated segment}
%    \label{fig:denoise_EMG}
%\end{figure}
%
%\begin{figure}
%    \centering
%    \includegraphics[width=0.97\linewidth, height = 0.4\linewidth]{figures/BRN_mixed.pdf}
%    \caption{Denoising performance comparison on the Mixed-noise segment}
%    \label{fig:denoise_mix}
%\end{figure}

%\begin{figure*}[t]
%     \centering
%    
%     \subfloat[EOG dataset\label{fig:sub1}]{
%         \includegraphics[width=0.31\linewidth]{figures/BRAA_EOG (2).pdf}
%     }
%     \hfill
%     \subfloat[EMG dataset\label{fig:sub2}]{
%         \includegraphics[width=0.31\linewidth]{figures/BRAA_emg.pdf}
%     }
%     \hfill
%     \subfloat[Mixed EOG/EMG dataset\label{fig:sub3}]{
%         \includegraphics[width=0.31\linewidth]{figures/BRAA_mixed.pdf}
%     }
%    
%     \caption{Routing visualization results on the EOG, EMG, and mixed EOG/EMG datasets.}
%     \label{fig:routing_vis_all}
%% \end{figure*}

\section{Results and Discussion}
To assess the effectiveness of the proposed \textit{BandRouteNet}, we compare it against several neural network-based EEG denoising methods, including the benchmark baselines FCNN~\cite{eegdenoisenet}, SimpleCNN~\cite{eegdenoisenet}, RNN-LSTM~\cite{eegdenoisenet}, and 1D-ResCNN~\cite{eegdenoisenet}, as well as recent advanced architectures such as EEGDNet~\cite{EEGDnet}, Deep Separator~\cite{deepseparator}, and LRR-Unet~\cite{lrr_unet}. To ensure a fair comparison, all competing methods are reproduced and evaluated under a unified experimental protocol, including the same  training, validation, and test splits, identical evaluation and visualization settings. 

\begin{figure*}[t]
	\centering
	
	\subfloat[EOG-contaminated EEG signal\label{fig:sub1}]{
		\includegraphics[width=0.31\linewidth]{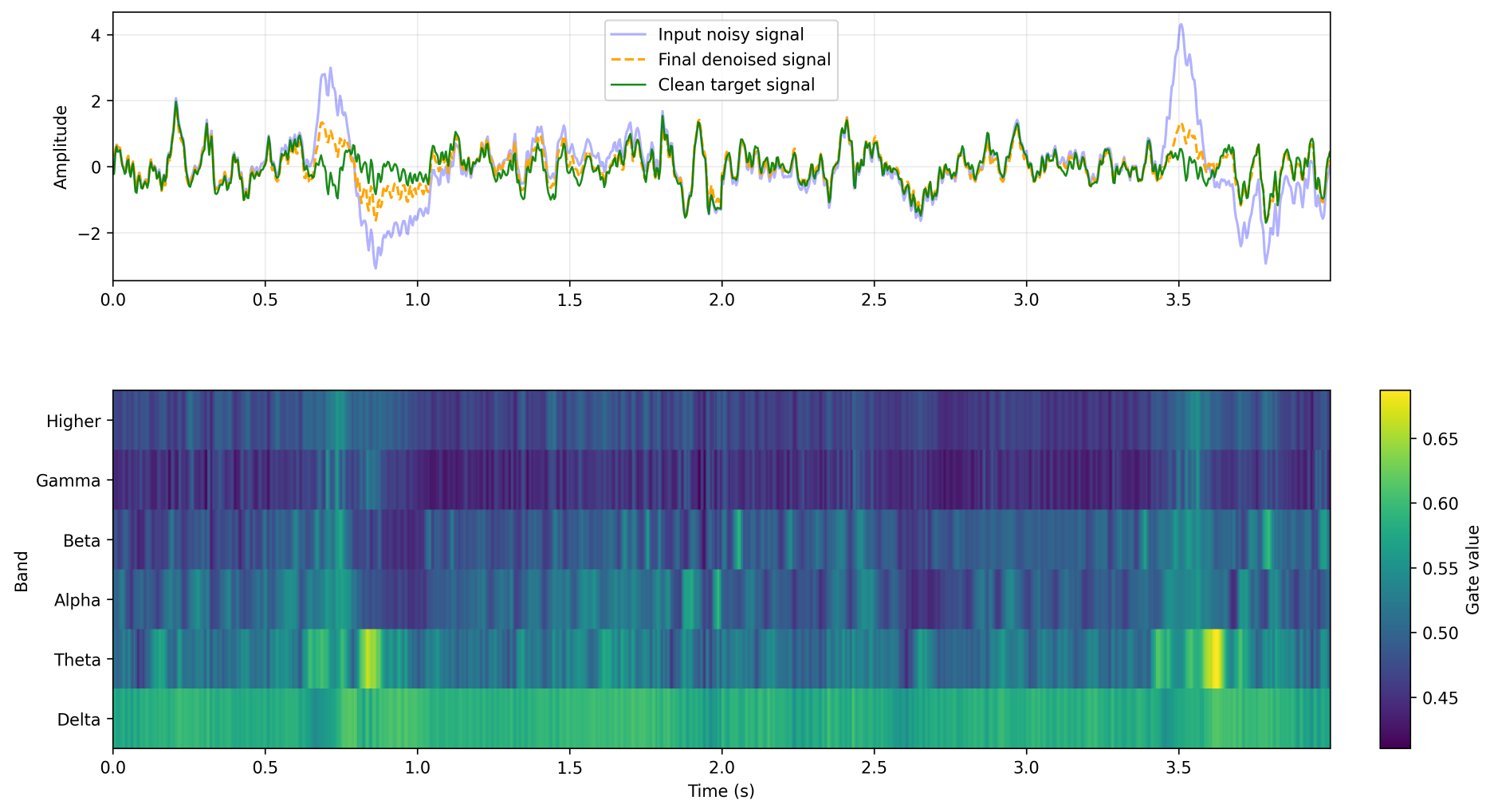}
	}
	\hfill
	\subfloat[EMG-contaminated EEG signal\label{fig:sub2}]{
		\includegraphics[width=0.31\linewidth]{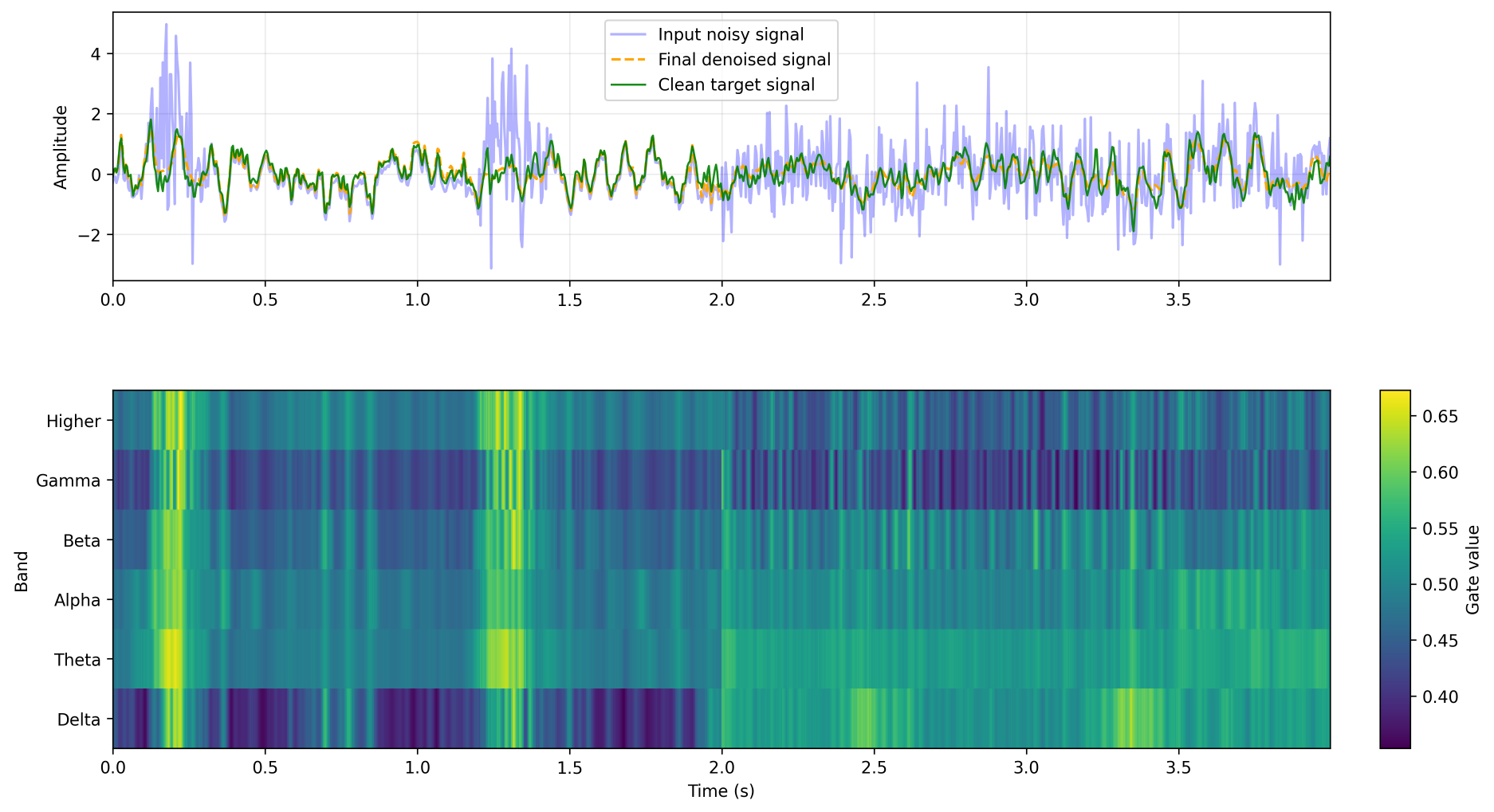}
	}
	\hfill
	\subfloat[EOG/EMG-contaminated EEG signal\label{fig:sub3}]{
		\includegraphics[width=0.31\linewidth]{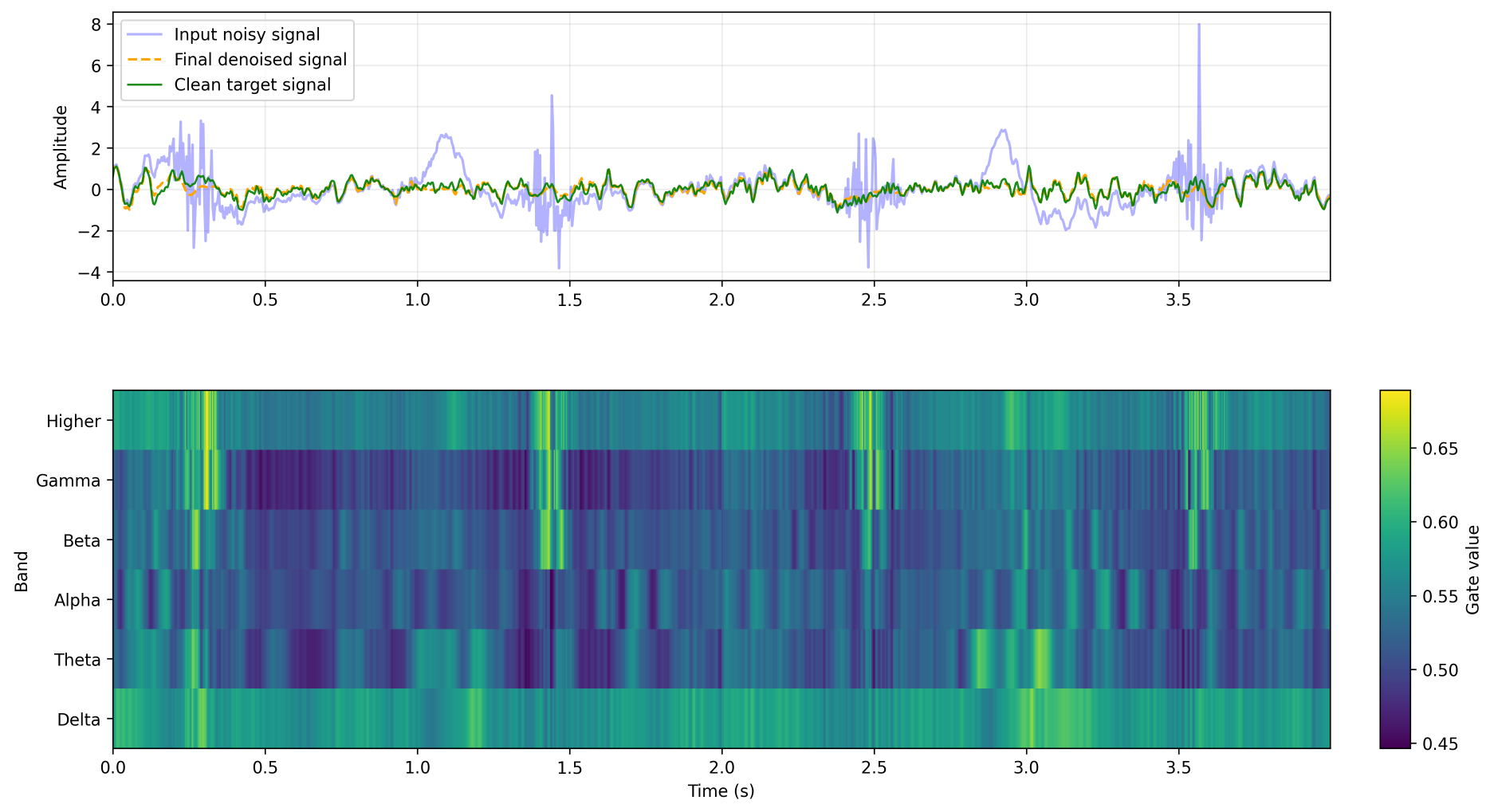}
	}
	
	\caption{Artifact routing visualization results under three noise conditions. In each sub-figure, the upper panels present the denoising results, while the lower panel shows the outputs of the Artifact Router module as a heatmap over time and frequency bands.}
	\label{fig:routing_vis_all}
	\vspace{-0.42cm}
\end{figure*}

\subsection{Qualitative Evaluation in the Time and Frequency Domains}
We first conduct visual analysis to compare denoising performances of all methods under EOG, EMG, and mixed-noise conditions using reconstructed waveforms and PSD curves, as shown in Fig.~\ref{fig:denoise_EOG}, Fig.~\ref{fig:denoise_EMG}, and Fig.~\ref{fig:denoise_mix}. Across all scenarios, \textit{BandRouteNet} produces reconstructions that more closely follow the clean target while effectively suppressing artifacts without introducing residual noise or over-smoothing. Its PSD curves also better match the clean spectra, indicating stronger preservation of EEG spectral characteristics. These results suggest that \textit{BandRouteNet} achieves a favorable balance between artifact removal and intrinsic EEG signal preservation.

\subsection{Quantitative Comparison Across Evaluation Metrics}

Table~\ref{tab:benchmark_results} reports the average denoising performance across all SNR levels on the EOG, EMG, and mixed EOG/EMG datasets. The proposed \textit{BandRouteNet} achieves the best overall results in all three settings. On the EOG dataset, it obtains the lowest $\mathrm{RRMSE}_t$ and $\mathrm{RRMSE}_s$ and the highest $\mathrm{CC}$ and $\mathrm{SNR}_{\mathrm{imp}}$, indicating improved waveform reconstruction, spectral preservation, and noise suppression. Similar improvements are observed on the EMG dataset, where \textit{BandRouteNet} achieves the best reconstruction errors and SNR improvement while remaining competitive in correlation. In the more challenging mixed-artifact setting, the proposed model again achieves the best performance across all metrics, demonstrating its robustness to heterogeneous artifact contamination.

The model performance on EOG artifact removal across different SNR levels are shown in Fig.~\ref{fig:eog-multi-snr}. While all methods generally improve as the input SNR increases, \textit{BandRouteNet} maintains consistently strong performance over the full SNR range. The advantage is particularly clear under low-SNR conditions, where artifacts are more severe and adaptive denoising is more critical, suggesting the robustness of the proposed frequency-aware routing mechanism across different noise intensities.

In addition to its denoising performance, \textit{BandRouteNet} is highly parameter-efficient. It contains only 0.20M trainable parameters, substantially fewer than most competing methods, while achieving superior average performance. This indicates that the proposed architecture provides an effective trade-off between denoising quality and model complexity.

\subsection{Ablation Analysis of the Proposed Network Components}

Table~\ref{tab:ablation} presents the ablation results on the EOG dataset. The full model achieves the best performance across all metrics, confirming the contribution of the proposed components. Removing the artifact routing gate causes the largest degradation, with $\mathrm{RRMSE}_t$ and $\mathrm{RRMSE}_s$ increasing from 0.3831 and 0.3797 to 0.4091 and 0.4013, respectively. This verifies the importance of adaptive routing for selectively controlling denoising strength over time. Removing the full-band conditioner also reduces performance, indicating that global temporal context and full-band guidance are beneficial for band-wise denoising. In addition, removing the cross-band fusion module or the band identity embedding leads to consistent performance drops, showing that both inter-band interaction and explicit band information help improve reconstruction quality. Overall, the ablation study demonstrates that each component contributes to the final model, with artifact routing being the most influential module.

\subsection{Visualization of the Artifact Routing Mechanism}

To analyze the proposed routing mechanism interpretably, we visualize the Artifact Router outputs under EOG, EMG, and mixed EOG/EMG noise conditions in Fig.~\ref{fig:routing_vis_all}. In each subfigure, the upper panel shows the denoising result, while the lower panel presents the corresponding router activation map. Larger heatmap values indicate stronger denoising emphasis on the corresponding band-specific features. The activations vary adaptively across both time and frequency. For EOG noise, strong responses mainly appear in low-frequency bands, especially Delta ($\delta$, 0--4 Hz) and Theta ($\theta$, 4--8 Hz), during intervals with prominent ocular artifacts. For EMG noise, the responses shift toward higher-frequency bands, consistent with the broadband, high-frequency nature of muscle artifacts. Under mixed noise, the routing map combines both patterns, showing that the model can respond to coexisting low- and high-frequency contamination. The activations are also temporally localized and smooth, suggesting that the router increases denoising mainly in corrupted regions while preserving cleaner signal segments.

\section{Conclusion}

This paper proposed \textit{BandRouteNet}, an adaptive neural network for EEG denoising. By combining band-specific denoising with routing-based adaptive refinement, complemented by full-band contextual conditioning, the proposed model effectively suppresses diverse artifacts and preserve the intrinsic structure of EEG signals. Extensive experiments under various noise conditions show that \textit{BandRouteNet} achieves superior performance across various standard evaluation metrics, while remaining highly parameter-efficient. 

\addtolength{\textheight}{-8cm}   % This command serves to balance the column lengths

%\section*{ACKNOWLEDGMENTS}
%\printbibliography
%\begingroup
%\footnotesize
%\setlength\bibitemsep{0pt}
%\setlength\bibparsep{0pt}
%\printbibliography
%\endgroup

\begingroup
\renewcommand*{\bibfont}{\small}
\setlength\bibitemsep{0pt}
\setlength\bibparsep{0pt}
\printbibliography
\endgroup
% \addtolength{\textheight}{-11cm}   % This command serves to balance the column lengths
%                                   % the textheight of the last page by a suitable amount.
%                                   % This command does not take effect until the next page
%                                   % so it should come on the page before the last. Make
%                                   % sure that you do not shorten the textheight too much.

% \begin{thebibliography}{99}
% % \bibliographystyle{IEEEbib}
% % \bibliography{refs}
% \bibitem{pubmed_ds}Dernoncourt, F. \& Lee, J. PubMed 200k RCT: a Dataset for Sequential Sentence Classification in Medical Abstracts. {\em Proceedings Of The Eighth International Joint Conference On Natural Language Processing}. pp. 308-313 (2017).
% \end{thebibliography}
\end{document}